%% file: main.tex
\documentclass[sigchi,screen,nonacm]{acmart}

\usepackage{balance}
\usepackage{soul}
\usepackage{subfigure}
\usepackage{enumitem}
\usepackage{graphicx}

\usepackage{multirow, makecell}

\newcommand{\eg}{{e.g.,}\xspace}
\newcommand{\ie}{{\it i.e.,}\xspace}

\usepackage[ruled]{algorithm2e} 

\begin{document}
    \title{A Large-scale Examination of "Socioeconomic" Fairness in Mobile Networks}
	
	%
	\author{Souneil Park}
	\affiliation{
	  \institution{Telefonica Research}
	  \city{Barcelona}
	  \country{Spain}
	}
	\email{souneil.park@telefonica.com}
	\author{Pavol Mulinka}
	\affiliation{
	  \institution{Centre Tecnol\`{o}gic de Telecomunicacions de Catalunya}
	  \city{Barcelona}
	  \country{Spain}}
	\email{pmulinka@cttc.es}
	\author{Diego Perino}
	\affiliation{
	  \institution{Telefonica Research}
	  \city{Barcelona}
	  \country{Spain}
	}
	\email{diego.perino@telefonica.com}
	%
	%
	%
	
	\begin{abstract}
		\input sections/abstract
	\end{abstract}
	
	%



	\maketitle
	\input sections/introduction

	\input sections/related
	\input sections/data
	\input sections/methodology
	\input sections/fairness_statusquo
	\input sections/fairness_threats
	\input sections/discussion

    \appendix
	\newpage
	\balance
	\bibliographystyle{ACM-Reference-Format}
	\bibliography{biblio.bib}
\end{document}

%% file: sections/abstract.tex
Internet access is a special resource of which needs has become universal across the public whereas the service is operated in the private sector. Mobile Network Operators (MNOs) put efforts for management, planning, and optimization; 
however, they do not link such activities to  \emph{socioeconomic fairness}. In this paper, we make a first step towards understanding the relation between socioeconomic status of customers and network performance, and investigate potential discrimination in network deployment and management. The scope of our study spans various aspects, including urban geography, network resource deployment, data consumption, and device distribution. A novel methodology that enables a \textit{geo-socioeconomic perspective} to mobile network is developed for the study. The results are based on an actual infrastructure in multiple cities, covering millions of users densely covering the socioeconomic scale. We report a thorough examination of the fairness status, its relationship with various structural factors, and potential class specific solutions. 

%% file: sections/introduction.tex
\section{Introduction}

The debate on net neutrality \cite{NetNeut_NYT}, often framed as ``internet as a utility'', highlights the conflicting public/private aspects of the Internet. Regardless of the position towards the issue, there is a common agreement that quality Internet access is crucial and its necessity is universal across all people. However, ensuring such a universal quality access is a challenging goal that requires constant investments, e.g., infrastructural planning, deployment, operation and maintenance. As most Internet service provides are private entities operating in a competitive market, their objective might not be perfectly aligned with the needs of the general public. Thus, questions could be raised about whether there is any unfairness and discrimination in the service quality. 

In this paper, we put forward the topic of \emph{socioeconomic fairness}, and report our examination on a major UK Mobile Network Operator (MNO).
\footnote{As a precise definition or a measure of fairness does not exist, we use this term in a broader sense.} 
We believe socioeconomic fairness will be a compelling issue in mobile network operation as mobile Internet access become a primary means of information access for the general public. Socioeconomic fairness is also closely related to the recent digital inclusion initiatives of many government bodies (e.g.,\cite{EU_digital_inclusion, US_digital_inclusion}), which emphasize quality Internet access to all people. Many MNOs also recognize the initiative and are making efforts accordingly (e.g., \cite{Telefonica_DI,Orange_DI,Verizon_DI}, for example, by extending coverage. While the initiative highlights the importance of eliminating marginalization and exclusion, detailed efforts so far have not deeply explored the socioeconomic dimension.
For example, the current practices of MNOs are agnostic from the socioeconomic status of the customers. The management and optimization practices are based on Key Performance Indicators (KPIs), such as coverage monitoring, and voice/data service metrics, of each radio sector\footnote{A radio sector is the minimum  unit of network deployment and to which the end users' devices connect for voice and data services}. Furthermore, this KPI driven network management may create socioeconomic discrimination as a side effect. For instance, if users or areas of a certain socioeconomic class generates more traffic, MNOs could in turn deploy more resources and this could potentially further incentivise users to generate more traffic. 

Exploring socioeconomic fairness involves multiple challenges, not only requiring a large-scale population but also detailed measurements at the level of individuals (e.g., byte consumption of every TCP flow of a user). In addition, the scope of an analysis needs to cover various aspects, including network resource deployment, data consumption behaviors, and urban geography. 

To the best of our knowledge, our work is the first to explore socioeconomic fairness in mobile networks. 
Our study thoroughly examines the fairness status based on millions of users densely covering the socioeconomic scale in multiple cities and their data spanning over multiple years.  We take benefit of the MNO's actual network infrastructure, and the UK census which measures the socioeconomic status at a very fine geospatial granularity. A novel methodology that combines these two datasets is developed. The methodology enables a geo-socioeconomic perspective to mobile networks, which can be replicated and  facilitate follow up works. 

In addition to understanding the current fairness status, we explore the underlying factors and provide various lessons about the relationships, e.g., with population density, sector deployment density, data consumption patterns, and device distribution. We further make efforts to generalize the findings to other MNO's through an open dataset. Based upon the lessons, possible approaches to better fairness are also discussed.

%% file: sections/related.tex
\section{Related Work}
\label{sec:related}

The broader literature on inequality has observed structural unfairness in the access to social capital or scarce resources, and its negative consequences to the society. Studies have revealed unfairness in various areas: access to health care, quality jobs, housing, education, public safety, and wider social networks \cite{10.2307/30036965,10.2307/2580416,Musterd2005,10.1086/657114,pmid17848457,doi:10.1162/003355397555361}, generally disfavouring the poor and minorities and escalating their social isolation. In parallel, diverse policy efforts and social programs are being made to level the playing-field (for example, through direct aids  \cite{scouts2020} and infrastructural investment \cite{osci2011}). 

As mobile Internet access has become a vital resource for a large population, we believe it is imperative to examine potential unfairness or discrimination. 
While many works are made on large scale mobile network performance analysis (\eg, user mobility  \cite{wally,secci-hot-mobility}, network KPIs and planning  \cite{hotornot,nika-hot}, network performance metrics  \cite{Falaki:imc,mcpa}, and user Quality-of-Experience  \cite{andrawww,qoedoctor}), the body of literature does not employ the socioeconomic perspective.
The literature on digital divide puts much emphasis on the topic. However, the studies primarily take the users' perspective \cite{pearce2013digital}, and assess the inequality in terms of technology access, skills, and benefits from the technology. The literature has explored digital divide depending on country, race, gender, and various sociodemographic factors \cite{jackson2008race, clayton2013limits,chircu2009perspective}. 

In contrast, we view the problem from the service providers' perspective. 
In fact, there has been a number of works from the mobile computing community that share a similar perspective of serving a wider public. Perino et al. \cite{perino2020experience} reports the effort of bringing affordable Internet access to under-served, remote communities. A recent work by Sen et al. looks into Free Basics, an initiative of Facebook to provide zero-rated web services in developing countries, and study the performance impairments \cite{sen2017inside}. Elmokashfi et al.\cite{elmokashfi2017adding} specifically focus on availability of mobile broadband, and examine the status across Norway. 


There is a related branch of works that show correlations between socioeconomic level of customers and traditional phone usages (\ie calls and SMSs). For example, \cite{enrique-cdr-girona,enrique-tool} show that socioeconomic status can be inferred with good precision using a large number of Call Detail Records (CDR) features for a 500,000 habitants city. Similarly, \cite{cote1,cote2} argue that, by targeting large regions of developing countries rather than individuals, good classification precision can be achieved with only a few properties  (\eg volume of outgoing calls). While we explore a different dimension from these works, our work also focuses on the data traffic, which dominates today's mobile network usage~\cite{ericsson-traffic} instead of CDRs. 

%% file: sections/data.tex
\section{Data}
\label{sec:methodology}

We use two main data sources: (i) passive data usage traces of the MNO; (ii) Index of Multiple Deprevation (IMD), a socioeconomic segmentation of the geographic spaces of the UK.

Our study covers three cities in the UK: London, Birmingham and Liverpool. The selection of the cities was made by taking the top 3 cities by population to achieve scale of the analysis. As our investigation involves city-specific aspects (e.g., sector deployment and urban geography), it is inherently challenging to extend it to many cities. 
For the main findings, we present the results for the three cities and examine their consistency. However, we often focus our presentation on London as the city hosts a more diverse population of different classes than the other two. 

\begin{figure*}[t]
\centering
\subfigure[London]{
	\includegraphics[width=0.3\linewidth]{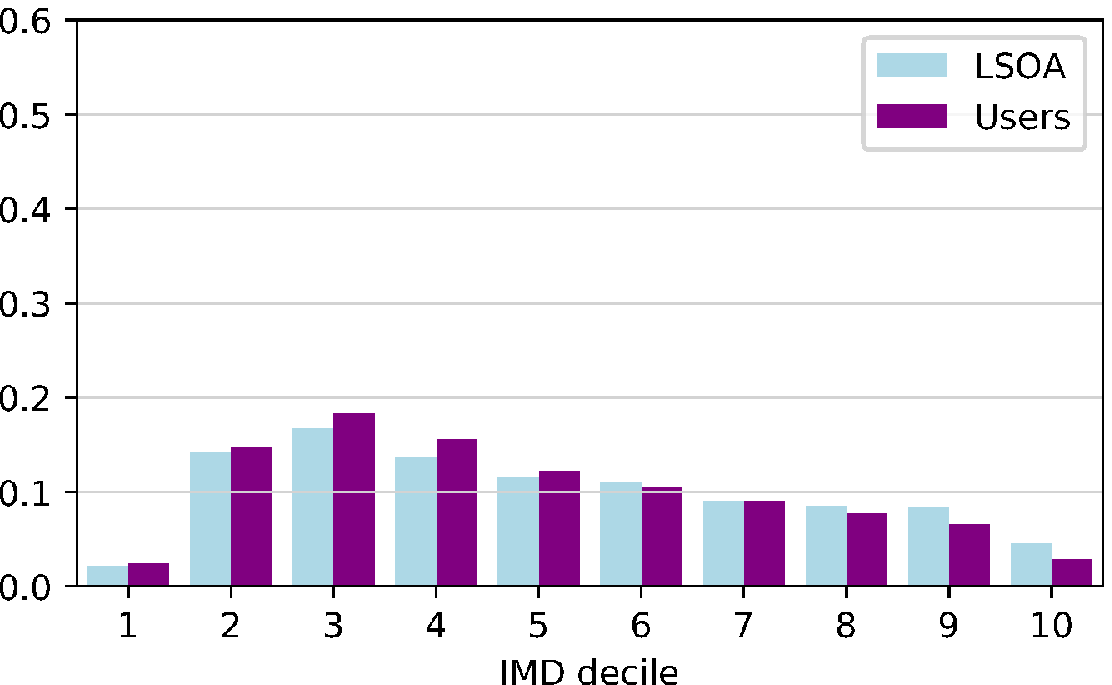}
	}
\subfigure[Birmingham]{
	\includegraphics[width=0.3\linewidth]{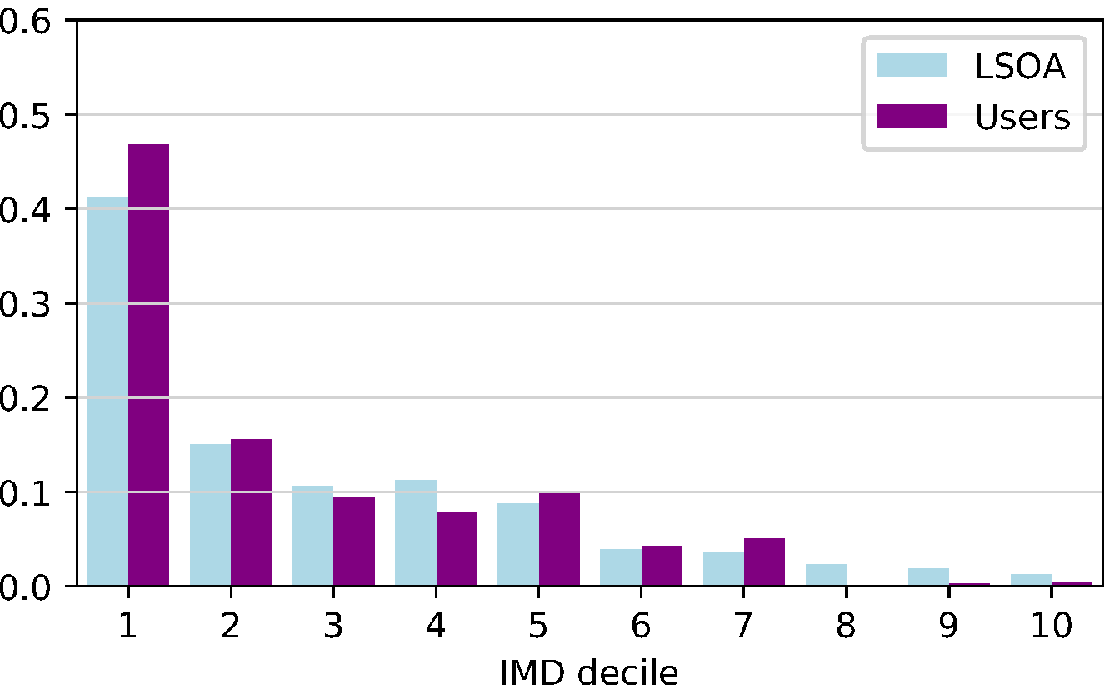}
	}
\subfigure[Liverpool]{
	\includegraphics[width=0.3\linewidth]{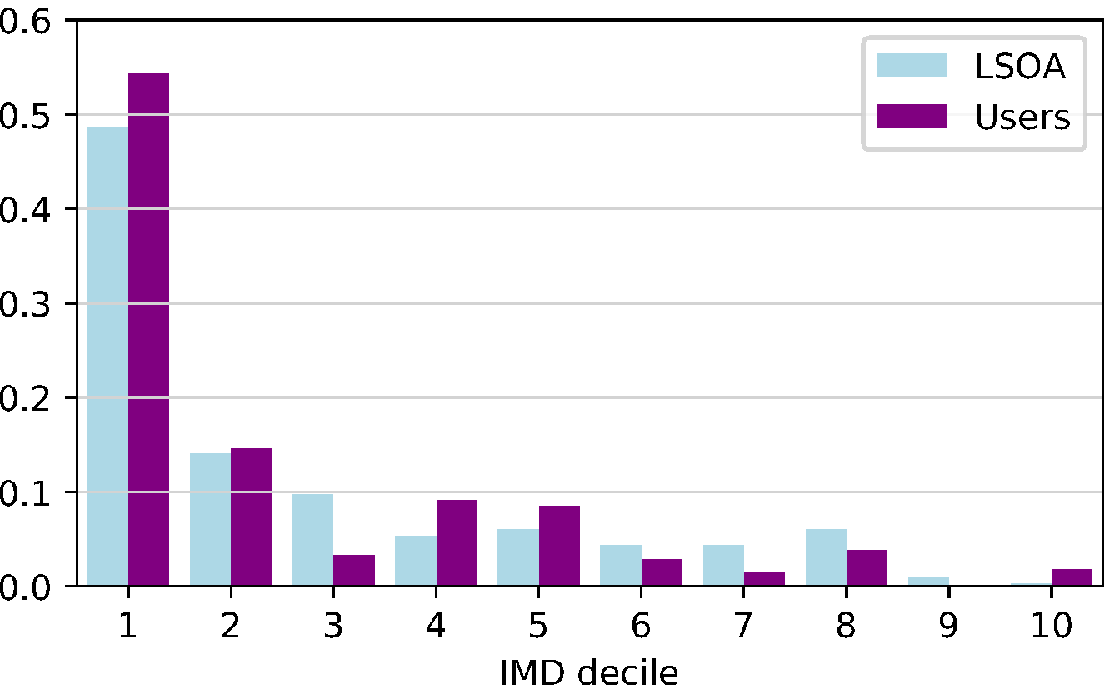}
	}
	\caption{Number of LSOAs per decile. (x-axis: IMD decile, y-axis: proportion).
	\label{fig:lsoa_cnt_decile}}
\end{figure*}

Our study is conducted over a three year time-span and the data was extracted from various periods. The main analysis of socioeconomic fairness uses the data from three periods: Oct.-Nov. of 2018, Feb.-Mar. of 2019 and Jan. of 2020. 
The results of all the periods are highly similar and the findings hold consistently. We thus often focus on the results from a particular year in the paper. 

In addition, we examine the consistency of the  findings in two more ways. First, we check seasonal consistency by sampling a month of each season from Spring, 2019 to Winter, 2020. Second, considering the deep impact of the COVID-19 pandemic, we add the corresponding period, February to Mid April of 2020 and re-examine the main findings. 

\subsection{Network performance logs}
We analyze data collected at a \emph{middlebox} used by the MNO to optimize traffic and to log performance metrics about each user's transactions. The dataset was filtered to include only mobile phones, not M2M or IoT devices. A  transaction is an entry in the monitoring logs generated by the middlebox, and corresponds to an individual flow handling encrypted or non-encrypted traffic generated by an app of a mobile phone. Specifically, for every flow, the middlebox stores the following layer four (L4) information: average \textit{Round Trip Time (RTT)}, number of \textit{packet retransmission}, and \textit{total bytes transferred}. 
The monitoring system then aggregates these metrics at the user-level (\ie, across all transactions of a given user). 

Note that latency and packet retransmission are among the key  metrics used to assess network quality. For instance, excessive RTT can hurt real time applications as streaming or VoIP, and generally impacts flow throughput. Also, a high number of packet re-transmissions are caused by network congestion, poor connectivity with weak signal or signal distortion. The middlebox is located at the PGw (packet network data gateway) /GGSN (GPRS gateway service node) in the packet core, hence the metrics reflect the performance within the MNO network (i.e., between the user device and the egress point towards the Internet). In order to view these performance measures through the lens of urban geography, we also analyze information stored at the \emph{Mobility Management Entity (MME)} that keeps track of the radio sector a mobile device is connected to and the coordinate of the sector. The information is used to identify the residence of users at a fine-grained level and their socioeconomic status.

We clarify that some results are normalized to avoid reporting raw values and respect the MNO dataset confidentiality policy (the details of the normalization method is explained with the results). In addition, the data used in the study does not include any personally identifiable information such as name or home address. 
We also emphasize that the performance measures of the study are independent from the type of contracts/plans since the
MNO we examine does not perform any manipulation of the network performance (\textit{e.g.}, throttling) by them. 
\subsection{Socioeconomic indicator: IMD}

IMD is a statistical measure maintained by the UK government. The measure quantifies the relative deprivation for the areas in England at a very fine spatial granularity. This index combines seven distinct dimensions of deprivation: (i) Income deprivation, (ii) Employment deprivation, (iii) Health deprivation and disability, (iv) Education, skills and training deprivation, (v) Crime, (vi) Barriers to housing and services, and (vii) Living environment deprivation. By combining the above seven dimensions, the overall IMD measure takes a wider interpretation of deprivation and looks into how much a group of people lacks of resources of various kinds, not just income. As such, the IMD measure  has been used in studies of various domains (e.g., \cite{rivas2017exposure, mcgillion2017randomised}).

The index is maintained for all areas of the spatial division of England, called Lower-layer Super Output Areas (LSOAs). LSOAs are small areas divided to have similar population size and social homogeneity. An LSOA has an average of approximately 1,500 residents or 650 households. There are 32,844 LSOAs in England, 4,835 in London, 639 in Birmingham, and 298 in Liverpool. IMD is updated around every 5 years; the last three were published in 2010, 2015, and 2019.

Based on the deprivation score, a ranking is made with all LSOAs in England. Beside the IMD scores, the government publishes also the \emph{deciles}, i.e., it splits the 32,844 LSOAs into 10 equal-sized groups, from the most deprived (index 1) to the least deprived (index 10). The LSOAs in decile 1 fall within the most deprived 10\% nationally and the LSOAs in decile 10 fall within the least deprived 10\%. The skyblue bars of Figure \ref{fig:lsoa_cnt_decile} shows the distribution of LSOAs of the three cities over the IMD deciles. 
While London is not dominated by a particular decile, the distribution of the other two cities is heavily skewed to decile 1. We discuss the impact of this skewed distribution throughout the paper.

%% file: sections/methodology.tex
\section{Socioeconomic Segmentation}

As the MNO does not identify the socioeconomic status of its users, we develop a method for a fine-grained socioeconomic segmentation. 
It first identifies the home sector of users, and then maps the IMD decile score of LSOAs to the sectors. 
\vspace{-.2cm}

\subsection{Home sector estimation}


While there are many methods for identifying home of people from mobility traces, they commonly aim at finding regularity during night time and/or weekends (\cite{kung2014exploring,bojic2015choosing}). Likewise, we develop a method for home sector detection specific to our dataset with the same intuition. 


We go through the MME data of all the days throughout a month for each device, and identify the ones who connects to the same sector at night time (midnight - 8AM) for a substantial number of days. 
To ensure that the device is staying in the same area during the night time window, we employ the radius of gyration to estimate the spatial deviation of a device and filter out the days when the estimated radius is larger than 2km. 
Finally, the method identifies the home sector of a device as the one which was connected at least two weeks during a month, and has the maximum connection duration. The devices that do not have a home sector that satisfies the criteria are excluded from the study. 
This method identifies the home for a large set of users: 2M for London, 0.24M for Birmingham, and 0.13M for Liverpool ($\approx$10-20\% of the population of each city) \footnote{https://www.statista.com/statistics/294645/population-of-selected-cities-in-united-kingdom-uk/}.


The purple bars of Figure \ref{fig:lsoa_cnt_decile} shows the distribution of the identified residents of the three cities. We believe the general similarity of the distributions to that of the LSOAs (i.e., Purple vs. Skyblue bars) supports the reliability of our sample and the home detection method. We also acknowledge the limitation of our data for Birmingham and Liverpool, whose population is heavily skewed to decile 1. As there are only a few LSOAs of higher deciles (e.g., above 7 in Birmingham), the user samples are very small for some deciles. As for decile 8 of Birmingham and decile 9 of Liverpool particularly, our dataset does not have a user sample, thus, we do not have results for them. 

\begin{figure}
\centering
	\includegraphics[width=.7\linewidth]{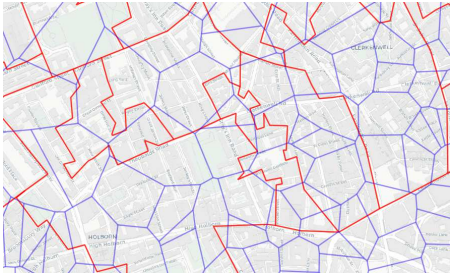}
	\caption{Boundaries of LSOAs (red) and Polygons (blue).
	\label{fig:lsoa_voronoi}}
\end{figure}

\subsection{Mapping IMD decile scores to sectors}
While the IMD measure is available by LSOAs, the mobility data only allows identifying the home of users by sectors. Thus, the IMD decile score needs to be estimated for the sectors in order to complete the socioeconomic segmentation. The challenge here is the mismatch between the boundaries of LSOAs and the coverage of the sectors. 

Depending on the deployment of the sectors, their coverage could be smaller than the boundary of one LSOA or it could span over multiple LSOAs. Therefore, we create a mapping between the sectors and the LSOAs by estimating the sectors' coverage and computing the overlap between the coverage and the LSOA boundaries. A Voronoi tesselation was used to estimate the coverage of antennas, which has previously produced reasonable results  \cite{lin2013towards, park2018mobinsight}. Subsequently, it was observed that the Voronoi polygons  frequently have a finer granularity than the LSOAs (The average size of the polygons is 0.33km$^2$ and that of LSOAs is 0.37km$^2$). 
Figure \ref{fig:lsoa_voronoi} depicts the boundaries of LSOAs and the Voronoi polygons for an example area in London. This suggests that it is feasible to estimate the IMD decile score of sector based on the LSOA it covers. In addition, both LSOAs and sectors are based on population density, such that small LSOAs are often associated with small Voronoi polygons.

\begin{figure*}[t]
\centering
\subfigure[Ratio of data consumed via 4G]{
	\includegraphics[width=0.31\linewidth]{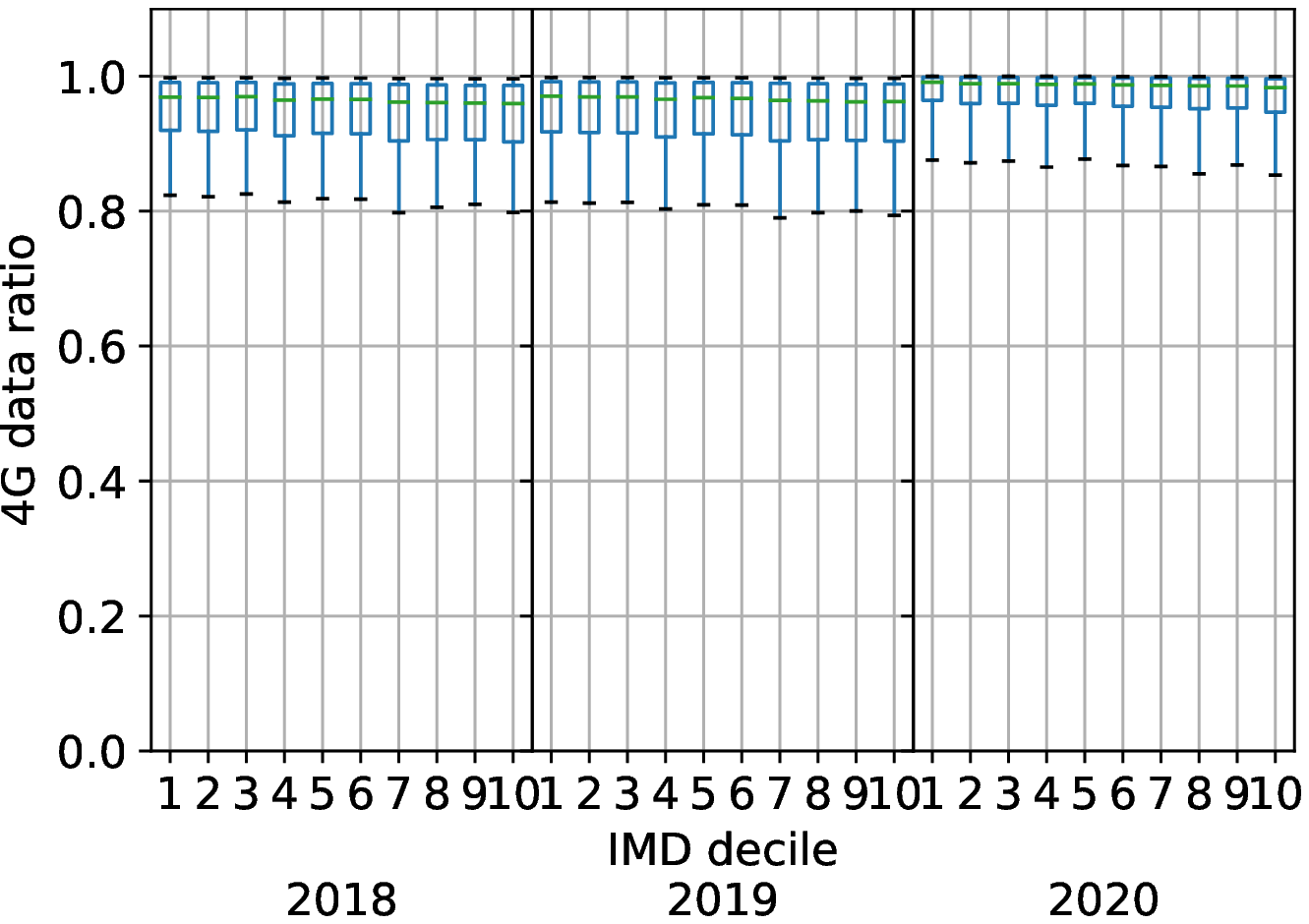}\label{fig:ratio4g-imd}
	}
\subfigure[Avg. packet retran. frequency (normalized)]{
	\includegraphics[width=0.31\linewidth]{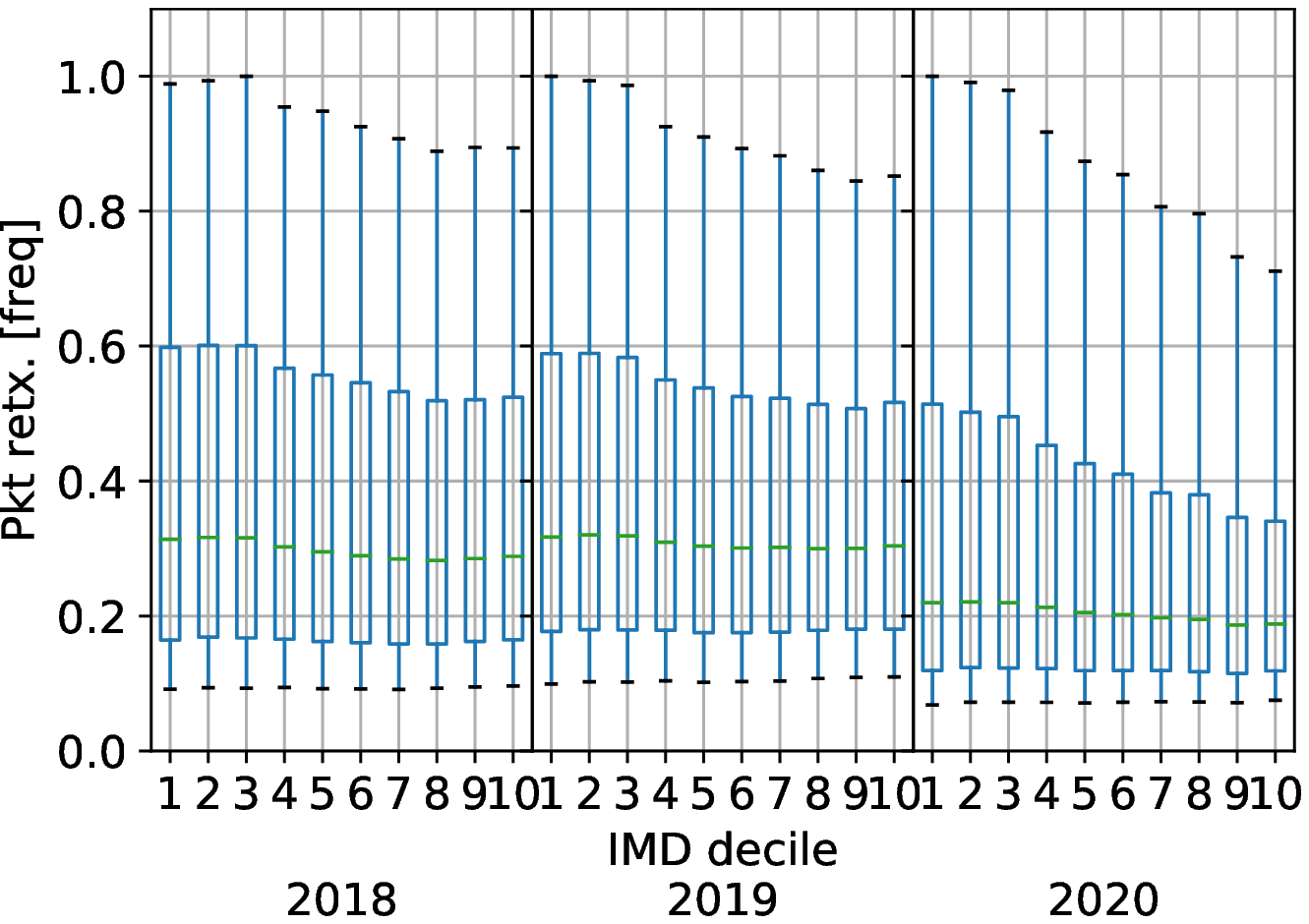}\label{fig:quality_user_imd_rtx}
	}
\subfigure[Avg. RTT (normalized)]{
	\includegraphics[width=0.31\linewidth]{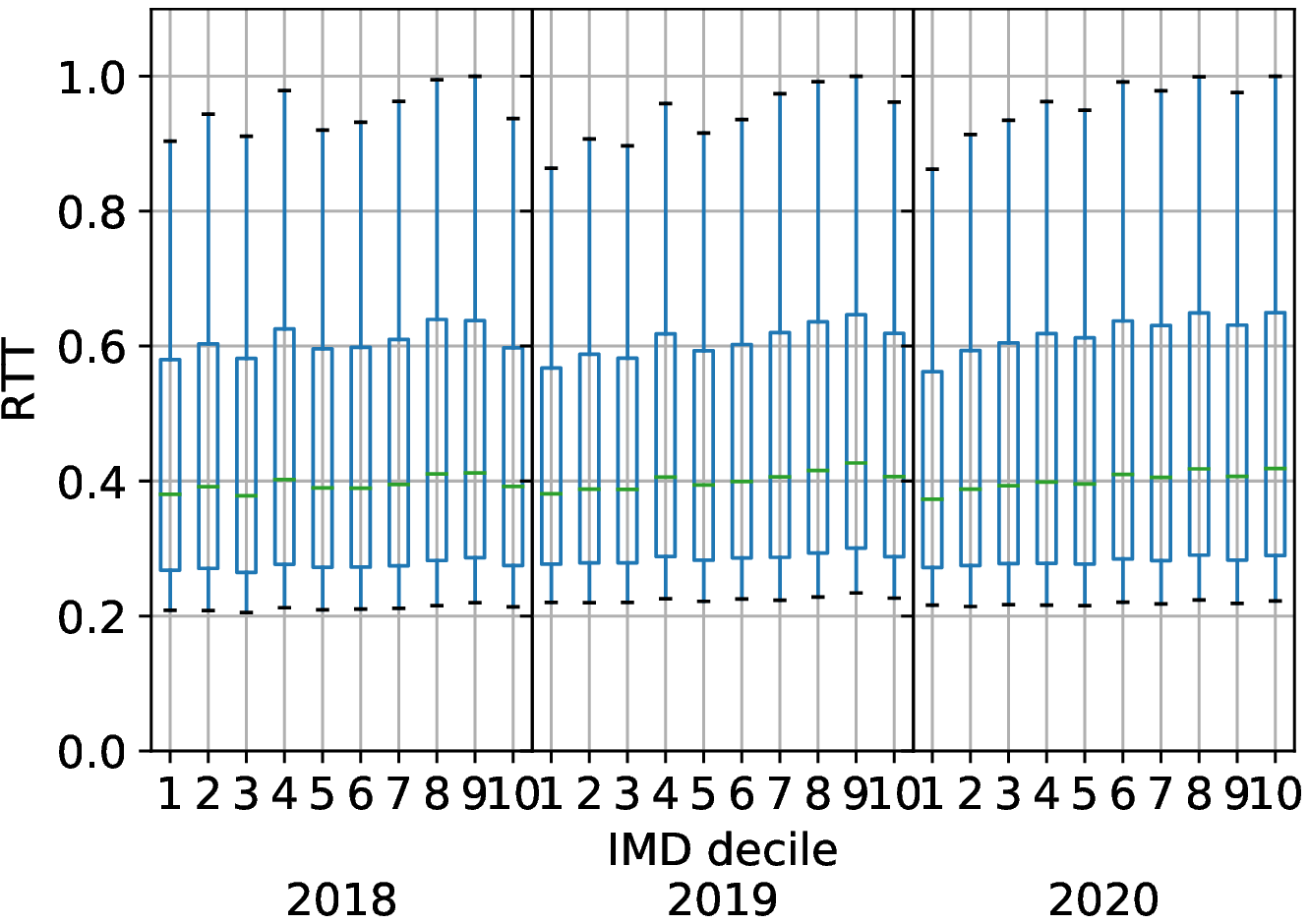}\label{fig:quality_user_imd_latency}
	}
	\caption{Network quality across socioeconomic classes and years in London\label{fig:avg_qual_ldn}}
\end{figure*}

\begin{figure*}[t]
\centering
\subfigure[Ratio of data consumed via 4G]{
	\includegraphics[width=0.31\linewidth]{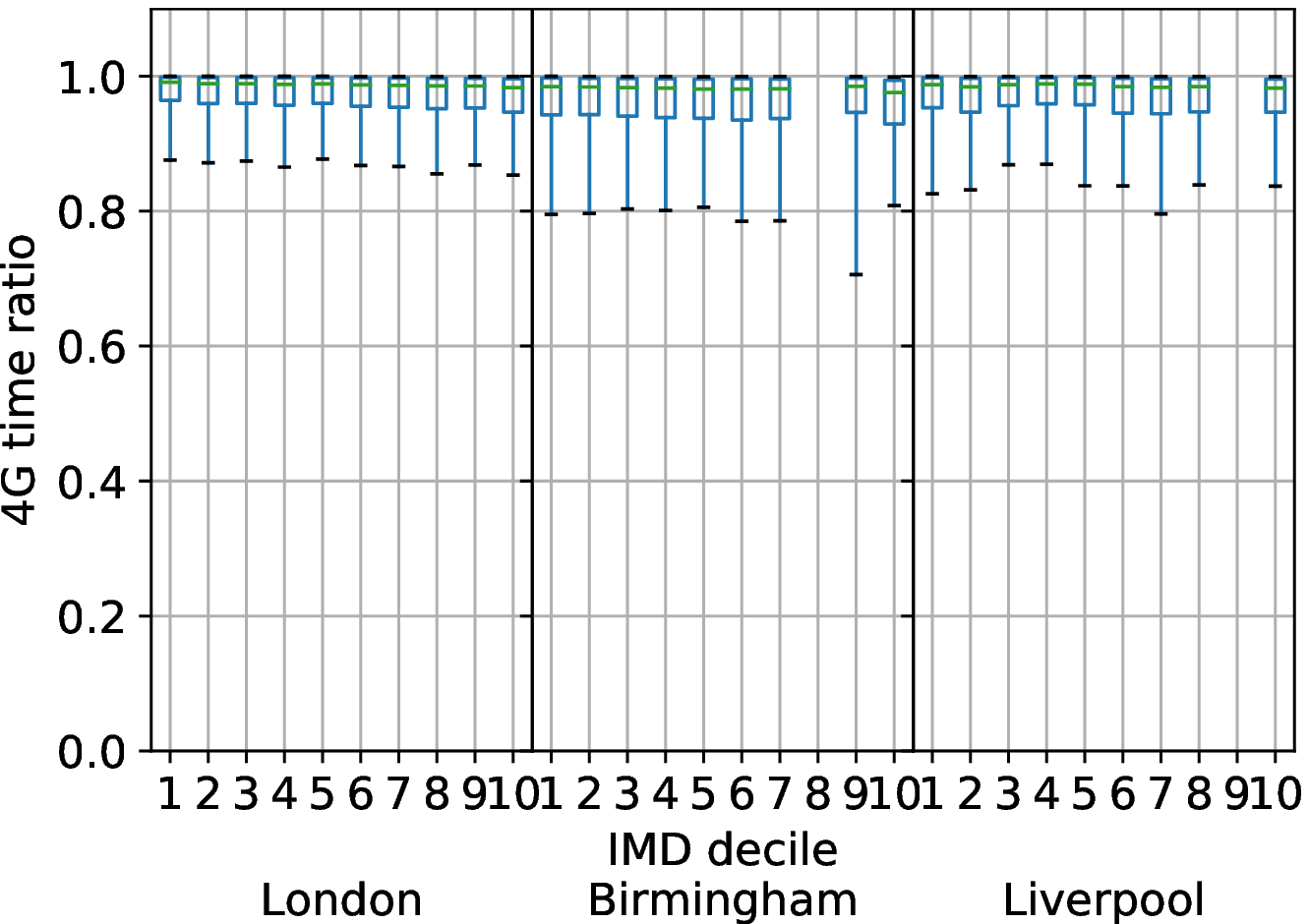}\label{fig:ratio4g_cities}
	}
\subfigure[Avg. packet retran. frequency (normalized)]{
	\includegraphics[width=0.31\linewidth]{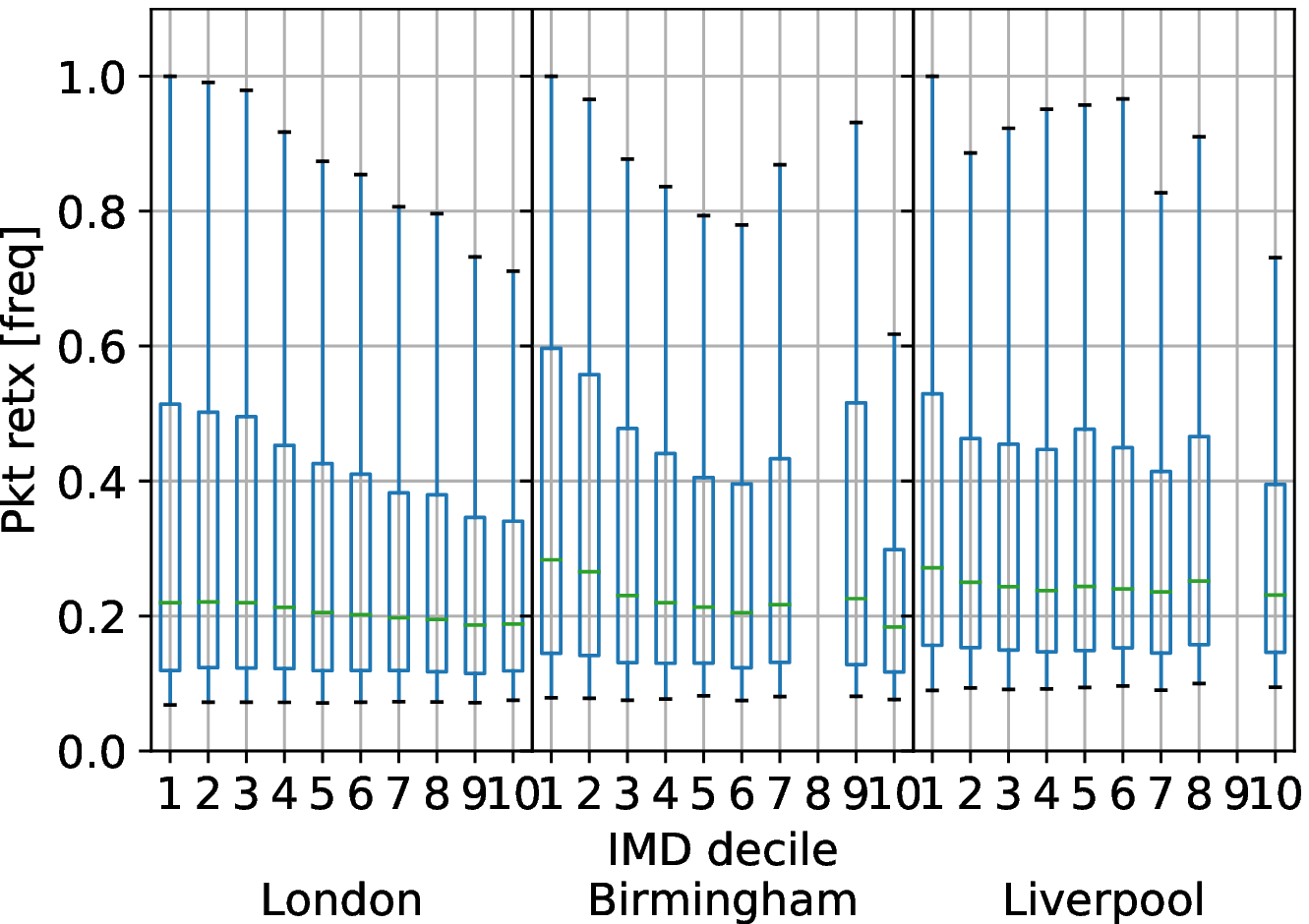}\label{fig:quality_rtx_cities}
	}
\subfigure[Avg. RTT (normalized)]{
	\includegraphics[width=0.31\linewidth]{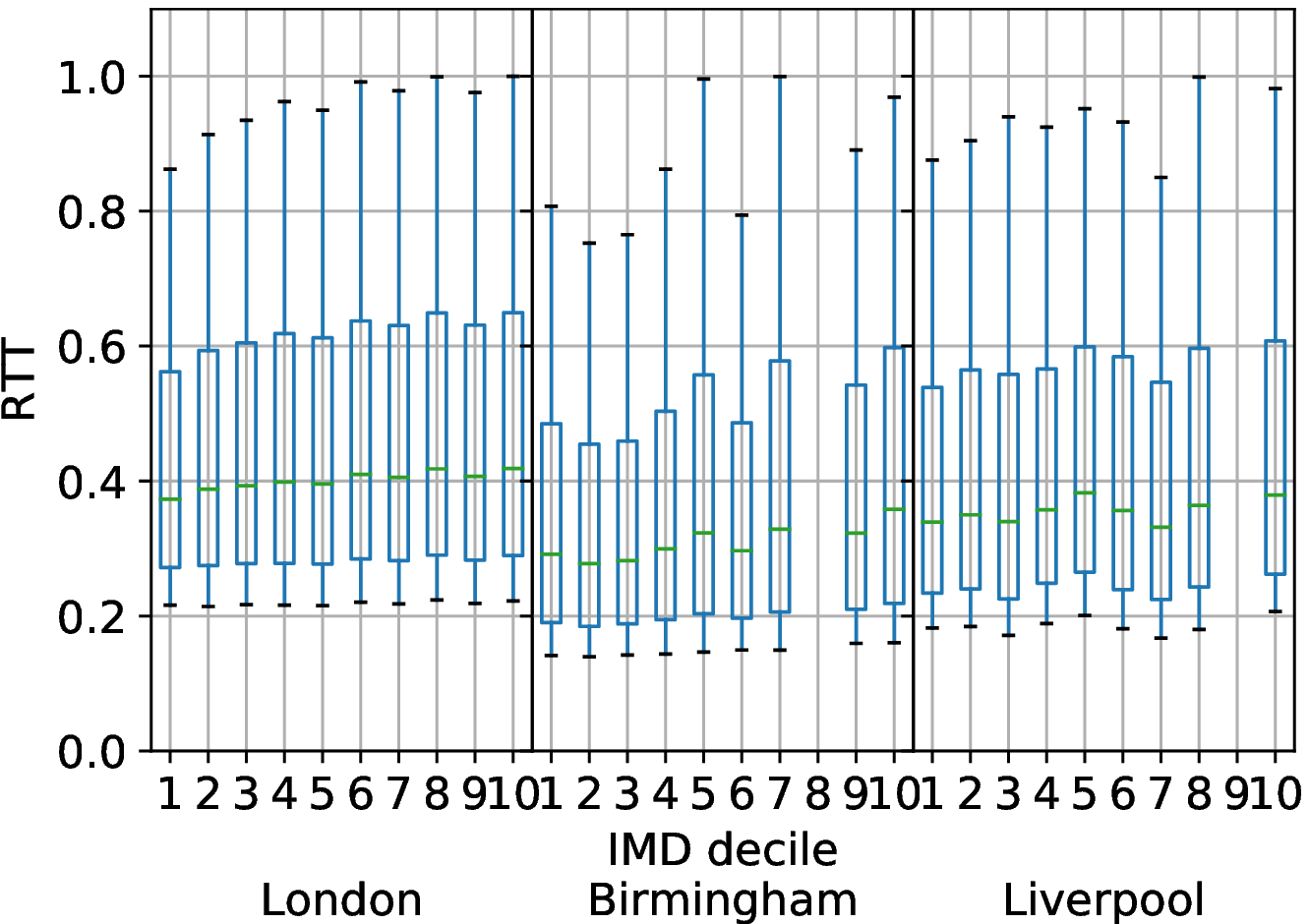}\label{fig:quality_latency_cities}
	}
	\caption{Network quality across socioeconomic classes and cities}
\end{figure*}

In some cases, a sector polygon overlaps with a single LSOA, making it simple to link the two units. However, many  polygons span multiple LSOAs (examples can be found in Figure \ref{fig:lsoa_voronoi}). In order to examine the impact of such cases on the study, it was first verified whether the LSOAs that overlap with a sector polygon are similar in terms of their IMD decile. For each of the sector polygons, the standard deviation of the IMD decile for the overlapping LSOAs was computed. It is observed that for the majority of the cases, LSOAs overlapping with a polygon tend to have similar levels of deprivation: for 72\% of the polygons of London, the standard deviation of the IMD decile scores is less than 1.5, 73\% for Birmingham, and 71\% for Liverpool. Thus the mapping is completed by simply taking the IMD decile value of the LSOA which has the largest overlap with a sector.

For the mapping of the IMD decile scores, we took the most recent IMD survey at the moment of the analyses. The IMD survey of 2015 was used for the analysis of the years 2018 and 2019. Although there is some time gap between the IMD survey and the analysis, we believe the mapping is valid since the IMD decile scores do not change much over time (e.g., the kendall rank correlation of the IMD deciles is 0.89 between the census of 2010 and 2015, and 0.84 between 2007 and 2015). The analysis of the 2020 data took the 2019 IMD survey, as it was updated in September 2019.

\vspace{-.2cm}

%% file: sections/fairness_statusquo.tex
\section{Overview of socioeconomic fairness}
\label{sec:fairness_sq}


To get an overview, we first segment the population of each city into 10 groups by IMD deciles and compare various quality measures described below. 


\noindent\textbf{Technology access}: As for the technology access, we break down mobile data consumption via technology, 4G LTE vs. others prior to 4G. 5G is not included in the measurement since it was not widely deployed during the study period, however, we discuss the implications to 5G deployment in the discussion section. 

For every user, we compute the ratio of data consumption made via 4G to that made via the others. Figure \ref{fig:ratio4g-imd} shows that, regardless of the socioeconomic status, most of the consumption is made via 4G. The median of the ratio is above 90\% for all classes, consistently across all the three years and cities. The variance within the deciles is also small, indicating that the 4G deployment covers most of the users and area. The figure also shows that the ratio further increases in 2020; the medians are slightly higher compared to the previous years for all the deciles, and the variance is also smaller.

\noindent\textbf{Network quality}: The quality is examined as a next step, focusing on 4G only as almost all consumption is made via 4G network. 
As discussed in Sec.~\ref{sec:methodology}, we employ two key metrics: average packet re-transmission frequency and average RTT. 
Fig~\ref{fig:quality_user_imd_rtx} compares the daily average packet re-transmission frequency across all deciles. For every user, this metric is computed by taking the ratio of packet re-transmitted over total number of packets of a flow, and then averaging them across all flows of a day. Fig.~\ref{fig:quality_user_imd_latency} compares the daily average latency (\ie average RTT experienced by the user, averaged across all flows of a day). Results are normalized over the highest 90-percentile to preserve MNO confidentiality. The results were simply re-scaled in order to preserve the trend in the data. The normalization was done separately for each year since our focus is inter-decile comparison within each year. Direct comparison between the years is not possible accordingly.

Figure \ref{fig:avg_qual_ldn} shows the results for London throughout the studied periods. The results do not reveal a clear evidence of unfairness, \ie at the level of IMD decile, the performance is not particularly better or worse among the 10 groups. 
Fig~\ref{fig:quality_user_imd_rtx} and Fig~\ref{fig:quality_user_imd_latency} show that the medians are very close and the central body of the distributions largely overlap among all the deciles. This is also consistent across the three years. 

We examine the two measures across the deciles in more detail by zooming into the data of one year, i.e., Feb.-Mar. of 2019. Two methods are used. First, we examine if a statistical difference is found among the deciles using ANOVA. While One-way ANOVA detects a significant difference, the significance is likely due to the scale of the samples, which is on the order of million. 
For example, one-way ANOVA confirms significant difference of re-transmission frequency among the deciles (F(9, 1024561) = 163.9, \textit{p} $<$ .01); however, the effect size (eta-squared) of this ANOVA model is .0016, which is far below the value (.01) that is considered as a general rule of thumb of a small effect size \cite{miles2001applying}. 
The result is similar for RTT as well, where the effect size is 0.0012.

Second, we also explore possible complicated associations that might not have been detected by ANOVA using a number of popular machine learning methods (\ie logistic regression, Gradient Boosted Trees, and SVM). The task is to identify the IMD decile of a user given the summary statistics of the user's network performance measures. While we omit the details, all three methods produce a result similar to a random classification, indicating that associations between network performance and IMD decile is hardly found.

In summary, the analysis made at the level of IMD deciles overall demonstrates negative results. However, visible trends appear in the upper tail of the distributions. Fig~\ref{fig:quality_user_imd_rtx} shows that the retransmission frequency gradually reduces along the IMD deciles whereas Fig~\ref{fig:quality_user_imd_latency} shows that the latency is slightly higher in some higher deciles. We investigate these upper tail trends in the next section.

\noindent\textbf{Network quality of other cities}:
The analysis on the other two cities confirms general consistency of the findings (recall that the data set does not have user samples for decile 8 of Birmingham and 9 of Liverpool, so the results are missing for them). Data consumption is mostly made via 4G for all deciles (Fig. \ref{fig:ratio4g_cities}), and the distributions of the packet retransmission frequency and RTT largely overlaps among all the deciles (Fig. \ref{fig:quality_rtx_cities} and Fig. \ref{fig:quality_latency_cities}). One-way ANOVA detects a significant difference among the deciles, but again only with weak or negligible effect sizes (eta-squared below .01). Interestingly, similarity to London is also found around the upper tail to some extent although the trend is a bit more noisy possibly due to the limited samples in the higher deciles. 

\noindent\textbf{Network quality across seasons \& COVID-19}: The plots are omitted to avoid redundancy as the same trends are observed.  One-way ANOVA detects a significant difference among the deciles, but again with negligible effect sizes (eta-squared below .01). During the pandemic period (Apr. 2020), the difference among the deciles became further negligible (eta-squared far below .01). 

\noindent\textbf{Network quality between weekdays and weekends}: For all the above analyses, we also separated weekends from weekdays and examined if notable difference of network quality is found. 
The results are omitted since all the trends described above were consistent between the two. However, we believe an interesting future direction is to further divide the hours at a finer-grained level and focus on peak hours. 


%% file: sections/fairness_threats.tex
\section{Socioeconomic Skews in Areas of Worse Performance}

\begin{figure*}[t]
\centering
\subfigure[Packet retransmission performance]{
	\includegraphics[width=0.51\linewidth]{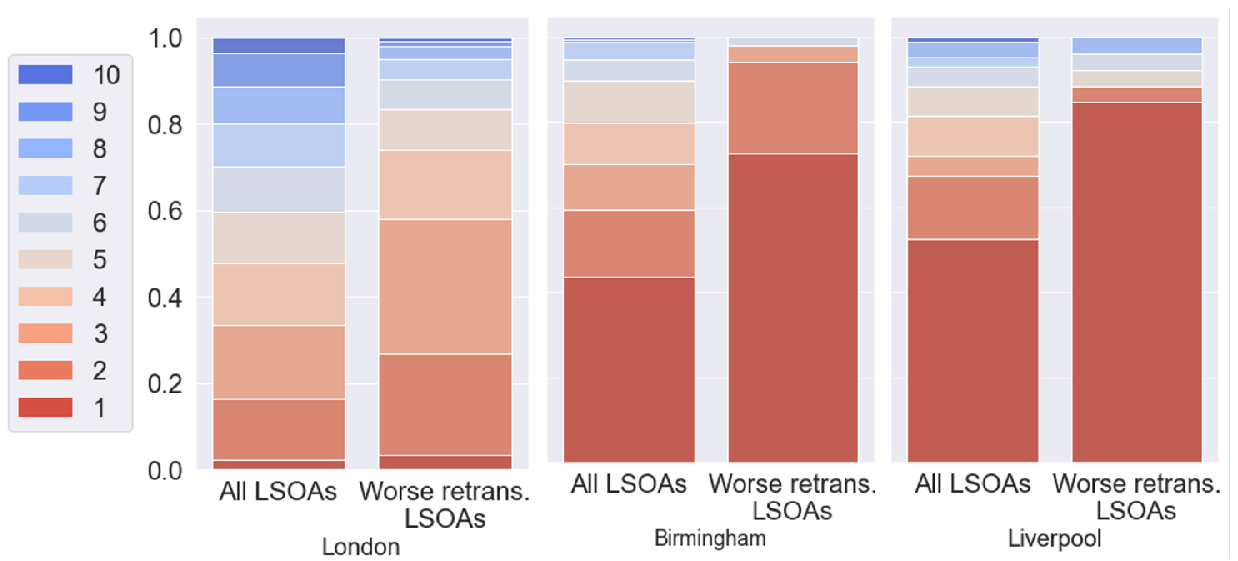}\label{fig:ut_pkt_ret}
	}
\subfigure[RTT performance]{
	\includegraphics[width=0.45\linewidth]{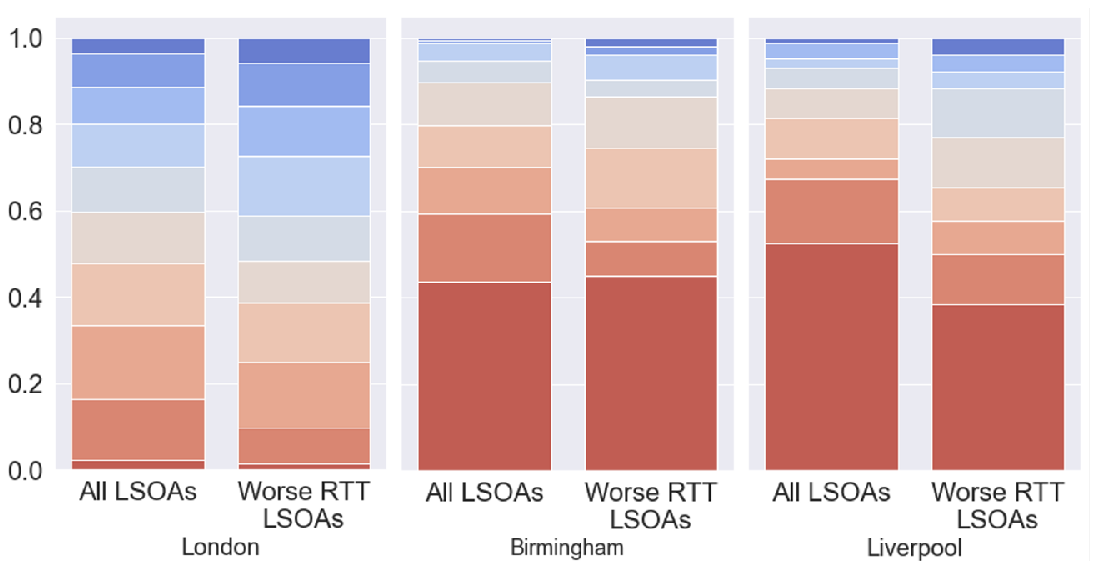}\label{fig:ut_rtt}
	}
\caption{IMD decile segmentation of LSOAs of worse performance. \label{fig:ut_IMD}}
\end{figure*}

In this section, we move beyond understanding the average performance and delve into the geographic areas of relatively worse performance, and examine socioeconomic bias among them. For a finer-grained view than the IMD decile segmentation, the population is further segmented by LSOAs. This finer-grained division allows exploring diverse factors that are potentially associated to the performance, including urban geography, sector deployment, and device type distribution. 

We aggregate the performance measures of individuals by LSOA and rank the LSOAs in the following two steps. First, for each of the two measures, we identify the set of users who fall below the third quartile (>75\%), whom we refer to as the inferior experience group. Second, the LSOAs are ranked by the ratio of residents who are included in the inferior experience group. We only present the results for the most recent period (Jan. 2020) of London since the findings hold consistently across the years and cities. 

Figure \ref{fig:ut_IMD} shows the IMD decile distribution for the LSOAs of worse performance, \ie the LSOAs with the highest (top 30\%) inferior experience group ratio. For comparison, the figure also shows the original IMD decile distribution of all the LSOAs of each city. We make two findings: first, the existence of socioeconomic skews in the areas of worse performance. For example, deprived LSOAs show a greater share among the areas of worse packet retransmission performance compared to their share in the general population. Second is the opposite trend between the performance measures. In contrast to the packet retransmission, less deprived LSOAs show a greater share among the areas of worse RTT performance. Both findings consistently hold for all three cities. The trends are also consistent with other LSOA percentile rank thresholds (\textit{e.g.}, 20\%, 10\%) for the highest inferior experience group ratio. 

The factors below are explored to understand the skews.

\begin{itemize}[leftmargin=*,topsep=0pt]
    \item user \& byte count: We count the number of residents and bytes consumed for every LSOA in order to approximate the load generated from the area.
    \item distance to sectors: This feature aims at comparing the density of the sectors deployed across LSOAs. However, the comparison is challenging since accurately estimating the coverage of the sectors and quantifying it by LSOA is difficult. Instead, for every LSOA, we measure the distance to \textit{k} nearest sectors from its centroid and take the average. 
    \item device type distribution: We identify and aggregate the device types of residents (i.e., manufacturer, brand, and model name) for every LSOA \footnote{We use a commercial database provided by GSMA to identify the device properties }. 
\end{itemize}

\noindent\textbf{Weak association to user/byte count}: A possible hypothesis for understanding worse performance is the relationship with the load of the LSOAs, \ie too many users or much data consumption affects the performance measures. However, we do not find support for the hypothesis. First, the correlation between the number of residents and the ranking of LSOAs is very weak for both performance measures (Table \ref{table:load_perf_corr}). 
Since correlation analyses are driven by general trends, 
we also specifically examine if the LSOAs of worse performance have a greater number of users or bytes than the others. A Mann-Whitney U test rejects the hypothesis. In fact, the test for both measures confirms that the median of the LSOAs of worse performance is lower than the median of other LSOAs.

\setlength{\textfloatsep}{-.3cm}
\begin{table}[t]
\footnotesize
\begin{tabular}[\linewidth]{|c|c|c|c|}
\hline
\multirowcell{2}{\textbf{User cnt.} - \\\textbf{Rank (retrans.)}} & 
\multirowcell{2}{\textbf{User cnt.} - \\\textbf{Rank (RTT)}} &
\multirowcell{2}{\textbf{Byte cnt.} - \\\textbf{Rank (retrans.)}} &
\multirowcell{2}{\textbf{Byte cnt.} - \\\textbf{Rank (RTT)}}
\\	
& & & \\ \hline
-0.128 & -0.028 & -0.024 & -0.152\\ \cline{1-4}
\hline
\end{tabular}
\caption{Correlation between user\&byte and LSOA ranking.\label{table:load_perf_corr}}
\end{table}

\noindent\textbf{Distance to sectors greater in LSOAs of worse RTT}: One could speculate that LSOAs with greater RTT may have sectors further away than others. However, a correlation analysis does not show a general trend that matches the speculation for both measures. We vary \textit{k} from 1 to 5 and compute the average distance to \textit{k}-closest sectors for all LSOAs. When \textit{k} is set to 5, a Spearman correlation analysis shows a weak correlation, 0.088, between the average distance and the RTT performance ranking of LSOAs. The coefficient is -0.210 for the average distance and the retransmission performance ranking. The values are similar for other \textit{k} values. 

As performance is affected by a variety of other factors than the distance to sectors, the absence of a clear trend is not surprising. However, when the question is framed specific to the LSOAs of worse RTT performance, we indeed confirm that the average distance is greater than others (Mann-Whiteney U value: 317897.0, \textit{p} < .01) with a substantial effect size (Common-language effect size of 0.582\footnote{Common language effect size is defined as ``the probability that a score sampled at random from one distribution will be greater than a score sampled from some other distribution." \cite{mcgraw1992common}}). We also examine the hypothesis  `LSOAs of frequent packet retransmission would show greater average distance' but do not find support.

\begin{figure}[t]
\centering
  \includegraphics[width=.9\linewidth]{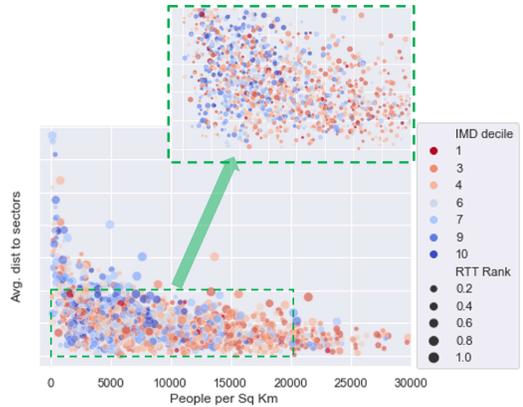}
\caption{Average Distance to Sectors by Population Density
\label{fig:dist_density_rtt}}
\vspace{.7cm}
\end{figure}

Projecting the average distance with respect the dimensions of population density and socioeconomic status provides insights for understanding the socioeconomic skew in the areas of worse RTT. In Figure \ref{fig:dist_density_rtt}, every LSOA is projected by their number of residents/$km^2$ according to the UK census and the average distance to sectors. The data points are colored by their IMD decile and sized by their RTT percentile rank (greater dots for worse RTT). We make three readings from the plot: first, the relationship between RTT and the distance to sectors, as bigger dots appear more frequently in the upper area of the plot. Second, sectors are sparser in the LSOAs of lower population density; the dots are centered in the left part as y-axis increases. Third, LSOAs of lower population density and sparser sectors are usually less deprived. Blue dots are increasingly frequent in the upper-left area. 

In addition to the population density, we also find that the amount of data consumption is associated to RTT. The number of bytes consumed is lower for the LSOAs of worse RTT than the rest (Mann-Whiteney U value: 215066.0, \textit{p} < .01) with a substantial effect size (Common-language effect size of 0.394). These observations altogether suggest that the socioeconomic skew against the less deprived LSOAs is likely linked to the population density and lower data demand rather that the socioeconomic status itself.
 


\noindent\textbf{Device type difference and packet transmission performance}: 
As we do not observe a clear association between packet retransmission frequency and the above infrastructural conditions, we extend the analysis to the devices of users. It is likely that the types of mobile devices vary between LSOAs of different socioeconomic conditions, and the devices could show different networking performance.

In order to explore the association, we compute the correlation between the ratio of residents with a particular device and the packet retransmission performance ranking of LSOAs. The correlation is computed for the 20-most popular devices in the city. As the distribution of devices follows a typical long tail distribution, looking into the 20-most popular devices covers 64.5\% of the whole devices in London.

We find a significant (\textit{p} < .01), moderate degree of correlation for a number of models; for example, the most popular device in London shows a Spearmann correlation coefficient of -0.346, and the second most popular device shows -0.348. The negative correlation between the share of a popular device and the LSOA ranking suggests that minor models that are more prone to worse performance could be more common in the LSOAs of greater packet retransmission frequency. As we look into the top-20 models, the correlation is negative in general: out of 15 models that show a significant correlation, 14 show a negative correlation. Among the 14 models, five show a moderate degree of negative correlation (<-0.25). We do not disclose the name of the models since the intention of the analysis is not to identify under-performing devices but rather to verify the existence of an association. 

We also run the same correlation analysis with respect to RTT; however, the correlation is unclear. The direction diverges (out of 18 models that show a significant coefficient, 9 are positive and 9 are negative), and the correlation is weak (all coefficients lie between -0.2 and 0.2). 

\section{Generalizability across Operators}

\begin{figure*}[t]
\centering
\includegraphics[width=.9\linewidth]{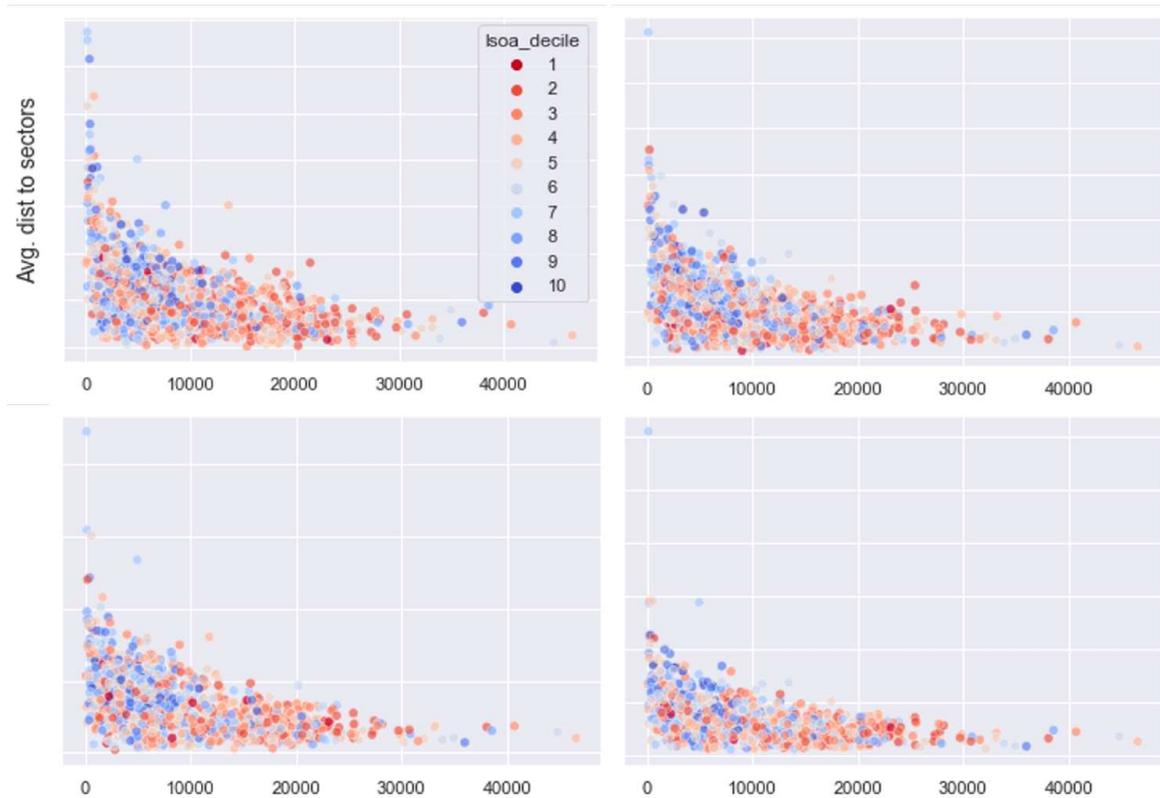}
\caption{Sector distribution overview of four MNOs (x-axis: average distance to 5-closest sectors, y-axis: population density)\label{fig:sector_other_mnos}}
\vspace{.7cm}
\end{figure*}

We make effort to ensure high validity of the findings, by covering a large sample of users, multiple cities, and multiple years. We also explore the possibility to extend the validity of the findings across other MNOs. 

Without having a similar type of performance measures from other MNOs, it is difficult to fully replicate the findings. However, we find the possibility of comparing sector deployment of multiple operators through open databases (\textit{e.g.}, OpenCellID). The comparison of sector deployment helps speculating on the socioeconomic skew in the LSOAs of worse RTT performance, assuming that the impact of the distance to sectors on RTT holds in general. 

We extract the sector deployment of three other operators in London from  OpenCellID (https://opencellid.org/). 
Figure \ref{fig:sector_other_mnos} plots the result of each operator in the same way used in Figure \ref{fig:dist_density_rtt} except sizing the dots by RTT ranking as the performance measure is missing. For comparison, we add a plot made with the data of the MNO studied in our work (upper-left). The same trends are observed, implying the possibility of the same socioeconomic skew against less deprived areas: sectors are sparser in the LSOAs of lower population density that are often less deprived. 

The consistency of the trends implies that the deployment practice could be similar between them. It also motivates future research for extending the validity of the findings we made with actual performance measures. The findings of our work could hold if the distribution of socioeconomic status of users, device types and data consumption behavior are not particularly different depending on the operators. 


%% file: sections/discussion.tex
\section{Lessons \& Conclusion}


In this section, we conclude the paper by summarizing the main findings and discussing the implications.

\noindent\textbf{Fair performance, on average}: 
The first finding is the balance of the average performance across the classes, which is relieving considering that our work was motivated to examine possible unfairness.
As discussed early on, this finding is compelling as mobile network has become a crucial resource for the general public, and understanding the fairness of it contributes to a broader range of research areas including social inequality and digital divide. In addition, the finding addresses a possible concern that deprived classes might be suffering from an infrastructural unfairness, which has been observed in many inequality studies. 
In fact, we see that the infrastructure management practice of operators, which in principle deploys more sectors in the areas of higher population density and data consumption, is leading to a denser deployment in deprived areas. 


\noindent\textbf{Two-sided socioeconomic skew}: However, our study does raise alarm for the LSOAs experiencing relatively worse performance as we see a socioeconomic skew among them. More importantly, we find that the socioeconomic skew is two-sided: each of the two performance measures we employ reveal opposing trends. Moreover, the two-sided socioeconomic skew is observed consistently across the three cities, which implies the existence of structural factors behind it. In a broader context, the finding provides implications to the research community and practitioners, highlighting the importance of considering urban geography in mobile environments and reflecting on subsequent socioeconomic impact. The opposing trends also call for future research on multiple different approaches towards addressing the skew. 

\noindent\textbf{Sparser sectors in less-deprived areas}: 
One side of the socioeconomic skew was associated to the sector deployment which could be seen as disfavouring the less-deprived class: less-deprived LSOAs more frequently show a sparser sector deployment and worse RTT performance. We further found relationships of sector sparsity with population density and data consumption, which shed light on possible solutions. For example, an MNO could pay special attention to LSOAs of lower population density and consider additional deployment even if the demand is relatively low. We believe the finding is timely as MNOs are progressively deploying 5G network: despite the difference of the technology from 4G, our work put forwards a new objective to consider in the deployment process, and provides lessons through a retrospective view of the 4G deployment about what geographical attributes could play a role and how. 


Related to the lessons, the interplay between sector deployment and diverse geo-socioeconomic factors is an interesting topic for future research. For example, demographic and lifestyle factors could be playing a role behind the data consumption behaviors. The portion of greater age population gradually increases along the IMD decile: people over 65 are 9\% in decile 1, and 19\% in decile 10; people over 45 are 29\% in decile 1, and 47\% in decile 10 \cite{lsoa_demographics}. Existing surveys have observed different degrees of engagement with technology, especially with smartphones, among different generations \cite{jiang2018millennials}. In addition, the relationship with different connectivity conditions (\textit{e.g.}, quality of home broadband and availability of WiFi access points) could be studied. A few prior articles already observed cases where people employ different channels for data consumption due to economic reasons \cite{bcg-report,ConnectivityDivide}. A deeper analysis of the traffic types (e.g., content types, apps, etc.) by areas is another direction of future work that will further expand the understanding of the performance dynamics. 

\noindent\textbf{Effect of devices than the infrastructure}: 
The analysis of the other socioeconomic skew we see for the packet transmission performance provides multiple interesting findings. First, we found an opposing trend where the deprived LSOAs were more prone to worse performance. Second, the skew did not show much association to the infrastructural conditions. Third, a moderate degree of correlation was observed between the share of popular devices and the performance. We believe our results provide empirical evidences that give guidance to future studies: whether to focus on the infrastructure or end-user devices, and particularly on which devices the focus should be. An immediate future work could conduct a large-scale examination of  end-devices and observe the performance impact more directly, particularly by including the devices that are more common in deprived LSOAs. Such a work would be important especially as 5G phones are being increasingly available in the market and adopted by people over time. The direction of research could also lead to new device benchmarks tailored to networking performance and initiatives for informing the public about the results, which can address the skew potentially.



\begin{acks}
This work has been partly supported by  the CHIST-ERA-17-BDSI-003 FIREMAN project funded by the Spanish National Foundation (under Grant PCI2019-103780), and the European Union’s Horizon 2020 research and innovation programme under grant agreement N° 101021808.
\end{acks}